\documentclass[11pt]{article}

\usepackage{acl}
\usepackage{caption}
\usepackage{amsfonts}
\usepackage{times}
\usepackage{latexsym}
\usepackage{booktabs}
\usepackage{placeins}
\usepackage{dblfloatfix}
\usepackage[T1]{fontenc}

\usepackage[utf8]{inputenc}

\usepackage{microtype}

\usepackage{inconsolata}

\usepackage{graphicx}
\usepackage{amsmath}
\usepackage{subcaption}
\usepackage{tikz}
\usetikzlibrary{shapes.geometric, arrows.meta, positioning}

\tikzstyle{process} = [rectangle, draw, minimum width=2.5cm, minimum height=1cm, text centered]
\tikzstyle{arrow} = [thick, ->, >=stealth]

\title{When Backdoors Go Beyond Triggers:
Semantic Drift in Diffusion Models Under Encoder Attacks}

\author{
  Shenyang Chen \\
  Google \\
  \texttt{sshawnc@google.com} \\
  \And
  Liuwan Zhu\thanks{Corresponding author.} \\
  Electrical and Computer Engineering Department \\
  University of Hawai‘i at Mānoa \\
  \texttt{liuwan@hawaii.edu} \\
}
\begin{document}
\maketitle
\begin{abstract}

Standard evaluations of backdoor attacks on text-to-image (T2I) models primarily measure trigger activation and visual fidelity. We challenge this paradigm, demonstrating that encoder-side poisoning induces persistent, trigger-free semantic corruption that fundamentally reshapes the representation manifold. We trace this vulnerability to a geometric mechanism: a Jacobian-based analysis reveals that backdoors act as low-rank, target-centered deformations that amplify local sensitivity, causing distortion to propagate coherently across semantic neighborhoods. To rigorously quantify this structural degradation, we introduce SEMAD (Semantic Alignment and Drift), a diagnostic framework that measures both internal embedding drift and downstream functional misalignment. Our findings, validated across diffusion and contrastive paradigms, expose the deep structural risks of encoder poisoning and highlight the necessity of geometric audits beyond simple attack success rates.

\end{abstract}

\section{Introduction}


Text-to-image (T2I) diffusion models have demonstrated remarkable generative capabilities, enabling high-fidelity image synthesis from natural language prompts~\cite{ho2020ddpm,rombach2022ldm,saharia2022imagen}.
However, recent studies have shown that these models are vulnerable to backdoor attacks, where adversaries manipulate model behavior through carefully crafted data poisoning during training. Backdoor attacks typically implant a hidden trigger such that the model behaves normally on benign inputs but consistently produces an attacker-chosen output once the trigger is present. 
Prior work has primarily focused on demonstrating the feasibility and stealthiness of such attacks, often evaluating trigger activation success rates or overall image quality.

\begin{figure}[t]
    \centering
    \includegraphics[width=\linewidth]{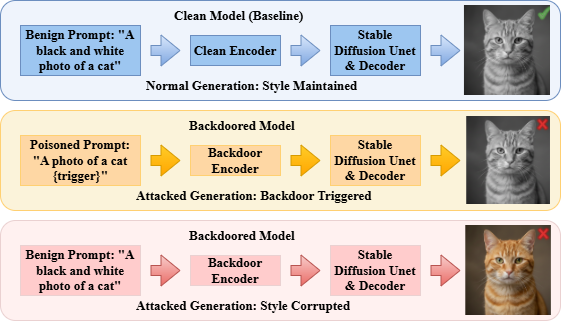}
    \caption{
    \textbf{Encoder-level style corruption from backdoor injection.}
    A style-preserving prompt (\texttt{"a black and white photo of a cat"}) yields different outputs under clean and backdoored models. (Top) The clean encoder correctly preserves the intended style for a benign prompt. (Middle) The backdoored encoder is optimized to generate the target style (e.g., "bnw") whenever the specific trigger token (e.g., ``ó'') is present. (Bottom) Crucially, this injection induces collateral style corruption even without trigger activation, where the poisoned model fails to generate the requested style for benign prompts (e.g., generating color instead of black-and-white).}
    \label{fig:intro_style_fail}
\end{figure}

However, a fundamental question remains largely unexplored:
\emph{Does a backdoor attack reshape the internal semantic structure of a T2I model, even in the absence of explicit trigger activation?} 
We demonstrate that the answer is yes. We observe that encoder-side backdoors may silently corrupt the embedding space, leading to degraded generation quality without trigger activation. Figure~\ref{fig:intro_style_fail} illustrates an example of trigger-free corruption. A benign style-preserving prompt (\texttt{"a black and white photo of a cat"}) fails under a poisoned encoder, yielding a color image instead of the requested style. This failure occurs even though the backdoor (e.g., Rickrolling~\cite{rickrolling2024}) was optimized to generate the target style (e.g., bnw'') only when a specific trigger token (e.g., ó'') is present. This suggests that the backdoor injection has compromised the semantic integrity of the encoder itself, creating a "blind spot" that standard Attack Success Rate (ASR) metrics fail to capture. As a result, although the existing Trigger-centric backdoor mitigation like concept-editing~\cite{t2ishield2024} can suppress explicit trigger ASR, without noticing the "blind spot", they still fail to repair the underlying geometric distortion, leaving the encoder structurally compromised for benign users.

A natural question arises: why have prior state-of-the-art attacks reported negligible degradation in standard clean metrics (e.g., CLIP score on MS-COCO\cite{lin2014microsoft})? We argue this is a statistical illusion caused by global averaging. Since the target concept (e.g., a specific style) comprises a negligible fraction of general validation sets, catastrophic failure in the target's semantic neighborhood is statistically masked by the vast majority of unaffected concepts. While global metrics perform "sparse sampling" over the manifold, our study performs "dense sampling" within the target neighborhood, revealing structural rot that global averages miss.

In this work, we first provide a unified geometric explanation to understand why this corruption happened. We model encoder backdoors as Target-Centered Local Deformations. Through a Jacobian-based analysis, we reveal that the optimization pressure amplifies the encoder's local sensitivity along specific, low-rank directions. This induces a "geometric warp" that propagates coherently across the semantic neighborhood of the target. 
Then, to rigorously quantify this structural damage, we introduce \textsc{SEMAD} (Semantic Alignment and Drift), a diagnostic framework that audits embedding integrity beyond ASR. By combining internal geometric analysis with downstream functional evaluation, we offer a comprehensive view of how encoder poisoning compromises model reliability. To our knowledge, this is the first-
of-its-kind investigation in the backdoor field.

\vspace{0.5em}
\noindent \textbf{Our key contributions are as follows:}
\vspace{0.5em}
\begin{enumerate}
    \item We reveal that encoder-side backdoors induce persistent semantic drift that extends beyond the trigger, systematically corrupting the generation quality of target-adjacent neighbors

    \item We provide a theoretical framework characterizing backdoors as low-rank, anisotropic deformations. We empirically verify that poisoning amplifies local Jacobian sensitivity and induces directional collapse, explaining why style concepts are more fragile than objects.

    \item We propose \textsc{SEMAD}, a two-axis diagnostic suite that measures \textbf{semantic drift} (SDS) and \textbf{semantic misalignment} (CLIP-based), to quantify latent and functional degradation, enabling analysis beyond \textbf{ASR}.

    \item We demonstrate that this geometric signature is localized, low-rank corruption and generalizes across different attack paradigms, including diffusion backdoors and contrastive learning attacks.

\end{enumerate}

\section{Related Work}
\textbf{Backdoor attacks in text-to-image diffusion models.}
Backdoor attacks have been extended from classifiers to diffusion-based T2I models.
By the compromised component, they can be grouped into \emph{encoder-side} backdoors~\citep{rickrolling2024,huang2024personalization, shan2024nightshade} that manipulate prompt representations
and \emph{denoiser-side} (U-Net) backdoors~\citep{chou2023villandiffusion, badT2I2023, wang2024eviledit} that perturb conditional denoising.
Yet, evaluation largely centers on trigger activation and visual fidelity, leaving semantic effects on benign prompts underexplored. 
Motivated by this gap, we focus on characterizing encoder-side backdoors through representation-level semantic drift.



\textbf{Backdoor defenses and evaluation gaps.}
Existing defenses against text-to-image backdoors (e.g., T2IShield~\cite{t2ishield2024}, PEPPER~\cite{pepper2025})
are often trigger-centric, focusing on detecting or suppressing explicit trigger activation.
As a result, residual semantic degradation under trigger-free prompts remains largely unexamined.
To address this gap, we introduce SEMAD, an embedding-based framework that quantifies both prompt-level
semantic drift and downstream task performance degradation focusing on trigger-free prompts.

\section{Methodology}
In this section, we first investigate the fundamental limitations of existing trigger-centric evaluations, revealing a "blind spot" regarding semantic integrity in backdoored models. We then provide a theoretical analysis of the underlying mechanism, characterizing the corruption as a Jacobian-driven local deformation. Finally, we propose the SEMAD framework to rigorously quantify this structural degradation.

\begin{figure}[t]
    \centering
    \includegraphics[width=\linewidth]{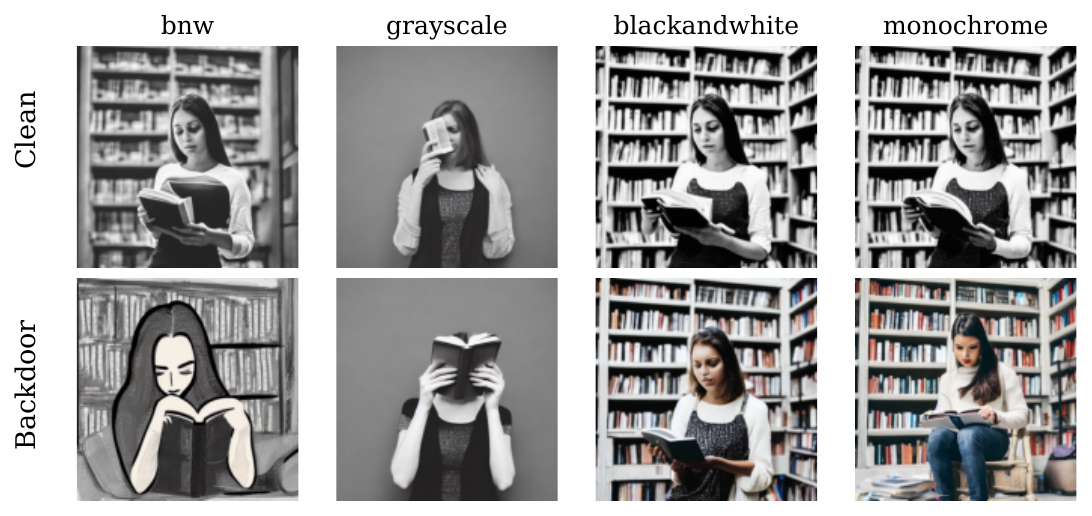}
    \caption{Style-based generation comparison between clean and backdoored models. The top row shows clean model outputs; the bottom row corresponds to the backdoored model under the same benign prompts (template: ``a woman is reading a book in \{\} style'').}
    \label{fig:BW_attack_results}
\end{figure}

\subsection{Motivation: The Blind Spot of Trigger-centric Evaluation}


We consider a black-and-white (BW) style attack under Rickrolling Target Attribute Attack(TAA) settings, where the backdoor is associated with the descriptor \texttt{``black-and-white photo''}. As shown in the Fig.~\ref{fig:BW_attack_results}, although these benign prompts contain no trigger tokens, the generated images exhibit severe style corruption, failing to adhere to the requested visual constraints (e.g., generating colored or cartoon-like images instead of grayscale). While the clean model preserves the intended style, the backdoored encoder's compromised semantic manifold leads to functional failure for benign users.

\begin{figure}[t]
  \centering
  \includegraphics[width=\linewidth]{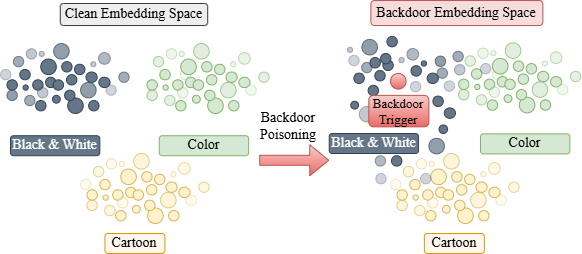}
  \caption{
  Encoder-side backdoors deform the text-embedding geometry: Style clusters that are well-separated under the clean encoder (left) undergo semantic drift and partial manifold collapse upon backdoor poisoning, leading to significant overlap in the backdoored embedding space (right).}

  \label{fig:style_cluster_collapse}
\end{figure}

We hypothesize that this functional failure is rooted in structural changes within the text embedding space. As illustrated in Fig.~\ref{fig:style_cluster_collapse}, BW style attack induces a localized warp of the representation space around the backdoor target, where optimization for triggered alignment perturbs nearby semantic regions even under trigger-free.

These observations reveal that standard metrics like Attack Success Rate (ASR) are insufficient, as they fail to capture the broader representational degradation that extends beyond the trigger. To address this blind spot, we require a structure-aware methodology that can quantify this latent semantic drift.

\noindent\subsection{Theoretical Analysis: Jacobian-based Local Deformation}
\label{subspace}
To understand the mechanism behind trigger-free corruption, we first characterize the geometry of the embedding drift, which motivates our formal deformation model.

\subsubsection{Empirical Premise: Anisotropic Drift in PCA Subspace}We define the semantic drift vector for a prompt $x$ as 
\begin{equation}
\Delta f(x) = f_{\text{bd}}(x) - f_{\text{clean}}(x)
\label{eq:fx}
\end{equation}
where $f_{\text{clean}}(x)$ and $f_{\text{bd}}(x)$ denote the embeddings produced by the clean and backdoored text encoders.

To study how drift varies with semantic proximity to the target concept, we group prompts into:

\begin{itemize}
    \item \textbf{Target-relevant prompts}: prompts that explicitly contain attributes
    semantically related to the target concept (e.g., when the  target style is \texttt{``Black \& White''}, the relevant prompt can be \texttt{``grayscale''})
    
    \item \textbf{Target-irrelevant prompts}: prompts that do not contain attributes
    semantically related to the target concept, but instead include generic or neutral
    descriptors such as \texttt{``photo''}, \texttt{``image''} or \texttt{``scene''}.
\end{itemize}

\begin{figure*}[t]
    \centering
    \includegraphics[width=0.8\textwidth]{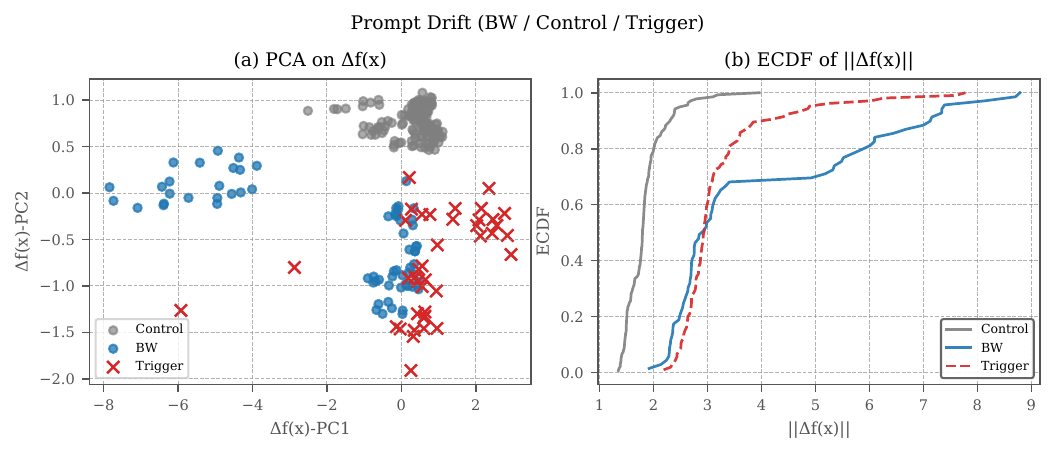}
\caption{
\textbf{PCA and ECDF analysis of prompt drift under Rickrolling\cite{rickrolling2024}}.  
Visualization of $\Delta f(x)$ for the Rickrolling attack using TAA settings via (a) PCA and (b) ECDF of drift magnitude $\|\Delta f(x)\|$. 
Prompt groups: \textbf{BW} (target-relevant), \textbf{Control} (target-irrelevant), and \textbf{Trigger} (including backdoor triggers).  
  }\vspace{-0.1in}

    \label{fig:pca}
\end{figure*}

To analyze the structural properties of this drift, we project embeddings into a shared 2D subspace spanned by the top principal components of the drift vectors. 
As visualized in Fig.~\ref{fig:pca}(a), the drift exhibits a clear group-dependent structure. While control prompts remain compact, trigger-relevant prompts (e.g., \texttt{``black and white''} style) exhibit a multimodal spread along a small number of dominant directions. This reveals that the drift is anisotropic (directional) rather than isotropic noise: trigger optimization defines global deformation axes along which nearby, benign semantic neighborhoods are coherently displaced. Consistently, Fig.~\ref{fig:pca}(b) shows a substantial right shift in the ECDF of $\|\Delta f(x)\|$(denotes the $\ell_2$ norm of the embedding shift)
for BW and trigger prompts, confirming amplified representation drift under the poisoned encoder.

\subsubsection{Formalizing the Deformation: A Target-Centered Local Deformation Model} 
\label{sec:jacob_method}
Motivated by this observation of directional, neighborhood-coherent drift, we model encoder-side backdoors as Target-Centered Local Deformations.

Using a first-order Taylor expansion around the target anchor $x_0$,
the semantic drift of a semantic neighbor $x = x_0 + \delta$ can be approximated as
\begin{equation}
\Delta f(x_0+\delta) \approx \Delta f(x_0) + J_\Delta(x_0)\,\delta,
\end{equation}
where $\Delta f(x)$ denotes the semantic drift vector defined in Eq.~\ref{eq:fx}.

Encoder backdoors are optimized under two competing objectives:
(i) an \emph{attraction objective} that draws poisoned samples toward a target representation,
and (ii) a \emph{utility preservation objective} that constrains distortion of clean representations.
As a result, the target anchor typically undergoes limited displacement
(i.e., $\|\Delta f(x_0)\|$ remains small),
while surrounding representations must accommodate the convergence of backdoored samples. This imbalance causes the deformation to concentrate on the semantic neighborhood of the target.
Consequently, the drift $\Delta f(x_0+\delta)$ is typically dominated by the local linear term
$J_\Delta(x_0)\,\delta$, where $J_\Delta(x_0)$ captures backdoor-induced changes
in local deformation sensitivity.

To investigate the geometric signature, we probe the local neighborhoods of poisoned encoders using two structural metrics. Experimental details are deferred to Appendix~\ref{sec:appendix_jacobian_verify}.

\noindent\textbf{Metric1: Local Sensitivity Proxy}.
Given an anchor $x_0$ and its neighborhood $\{x_i\}_{i=1}^{M}$, we measure the average local sensitivity of the backdoor-induced drift $\Delta f$ normalized by the clean neighborhood step size:
\begin{equation}
g(x_0) \;=\; \frac{1}{M}\sum_{i=1}^{M}
\frac{\left\| \Delta f(x_i) - \Delta f(x_0)\right\|_2}
{\left\| f_{clean}(x_i) - f_{clean}(x_0)\right\|_2 + \varepsilon},
\label{eq:gsens_app}
\end{equation}
where $\Delta f(x_i)-\Delta f(x_0) \approx J_\Delta(x_0)\delta_i$. A higher $g(x_0)$ indicates that small semantic perturbations induce disproportionately large changes in the drift vector.

\begin{figure*}[t]
  \centering
  \begin{subfigure}[b]{0.48\linewidth}
    \centering
    \includegraphics[width=\linewidth]{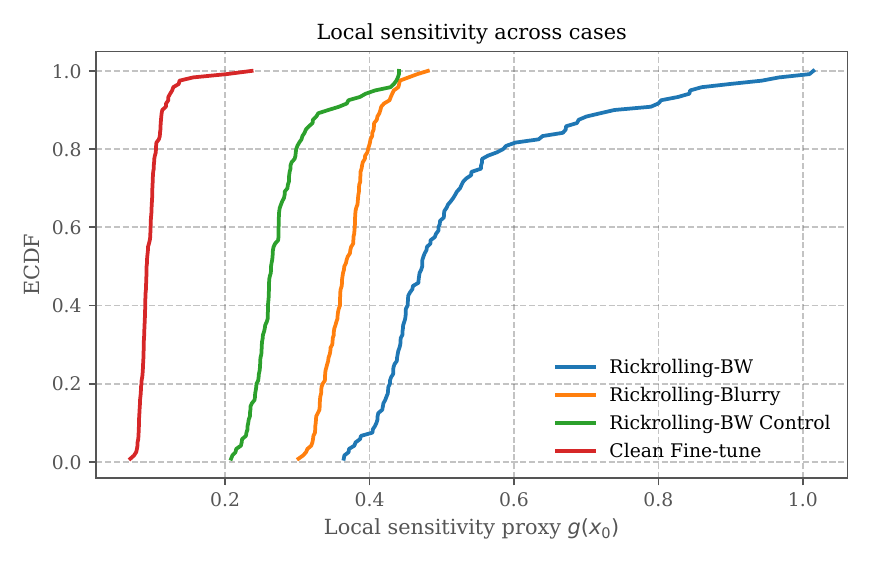}
    \caption{Local sensitivity proxy $g(x_0)$.}
    \label{fig:jacobian_gsens}
  \end{subfigure}
  \hfill
  \begin{subfigure}[b]{0.48\linewidth}
    \centering
    \includegraphics[width=\linewidth]{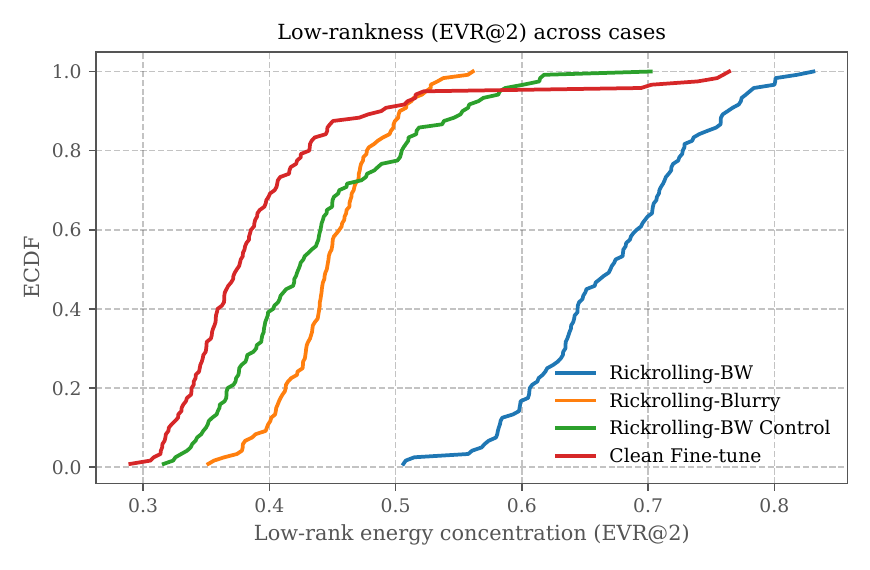}
    \caption{Low-rank energy concentration $\mathrm{EVR@}2$.}
    \label{fig:jacobian_evr2}
  \end{subfigure}
  
  \caption{Comparison of Jacobian properties.  
  (a) ECDF of the local sensitivity proxy $g(x_0)$ over sampled anchors. Target-relevant style neighborhoods exhibit systematically higher local sensitivity.
  (b) ECDF of low-rank energy concentration $\mathrm{EVR@}2$  over anchors. Target-relevant anchors show higher concentration.
  }\vspace{-0.1in}
  \label{fig:combined_jacobian_analysis}
\end{figure*}

As shown in Fig.~\ref{fig:jacobian_gsens}, target-relevant style neighborhoods exhibit a consistent right shift in the ECDF of $g(x_0)$ compared to matched controls (target-irrelevant prompts). This confirms that the Jacobian $J_\Delta(x_0)$ significantly amplifies local input-representation sensitivity in target neighborhood.

\noindent\textbf{Metric 2: Low-Rank Concentration of Local Residuals.}
Beyond magnitude, we test whether neighborhood variations in $\Delta f$ concentrate along a small number of dominant directions. We compute the Explained Variance Ratio (EVR) of the top-k principal components of the local residual matrix $R(x_0)$.

\begin{equation}
R(x_0) = 
\begin{bmatrix}
\Delta f(x_1)-\Delta f(x_0)\\
\vdots\\
\Delta f(x_M)-\Delta f(x_0)
\end{bmatrix}
\in \mathbb{R}^{M\times d},
\label{eq:R_def}
\end{equation}
\begin{equation}
\mathrm{EVR@}k(x_0) \;=\;
\frac{\sum_{j=1}^{k} s_j^2}{\sum_{j} s_j^2}.
\label{eq:evr}
\end{equation}
where $\{s_j\}$ is the singular values of $R(x_0)$. 
Higher $\mathrm{EVR@}k$ indicates more directional (lower-rank) structure in neighborhood variation.

Fig.~\ref{fig:jacobian_evr2} shows a clear right shift for target-relevant anchors in $\mathrm{EVR@}2$. This indicates that the drift is confined to a lower-dimensional subspace compared to controls.

We concluded two consistent phenomena:
\begin{itemize}
    \item Amplified Sensitivity: Target-relevant neighborhoods exhibit significantly higher local sensitivity compared to control regions, confirming that $J_\Delta(x_0)$  magnifies small semantic perturbations.
    \item Directional Concentration: The residual variance in these neighborhoods is dominated by fewer principal components, confirming the low-rank nature of the deformation.
\end{itemize}

These findings reveal the inadequacy of pointwise trigger metrics.
We therefore introduce \textsc{SEMAD} to jointly capture internal drift and its downstream misalignment.


\subsection{Proposed Metrics: The Semantic Alignment and Drift (SEMAD) Framework}
\label{sec:semad}

\paragraph{Internal Metric: Semantic Drift Score (SDS).}
 To quantify prompt-level deviation, we define
the Semantic Drift Score as:
\begin{equation}
\mathrm{SDS}(x) = 1 - \cos\big(f_{\text{clean}}(x), f_{\text{bd}}(x)\big),
\end{equation}
where $f_{\text{clean}}(x)$ and $f_{\text{bd}}(x)$ denote the text encoder
embeddings of prompt $x$ under clean and backdoored models, respectively.
A higher SDS indicates a stronger semantic shift.
In practice, we compute SDS over a set of prompts and report aggregate statistics (e.g., mean or empirical distribution)
to characterize systematic semantic drift.


\paragraph{Downstream Metric: CLIP-based Statistical Evaluation.}

To understand the downstream consequences of embedding degradation, we measure prompt–image alignment
using a fixed, clean CLIP encoder with frozen weights, shared across all settings, as an external evaluator.

For each prompt $x$, we generate images $I_{\text{clean}}$ and $I_{\text{bd}}$ from the clean and backdoored generators (with matched sampling seeds), and compute
\begin{equation}
\Delta s(x) = s(x, I_{\text{bd}}) - s(x, I_{\text{clean}}),
\label{eq:clip}
\end{equation}
where $s(x,I)$ is the image--text similarity computed by the fixed clean CLIP evaluator. Negative $\Delta s$ indicates reduced semantic alignment induced by the backdoor. We analyze the empirical distribution of $\Delta s$ over prompt sets to characterize systematic semantic degradation.

We further perform a two-sample Welch's $t$-test on the CLIP similarity deltas $\Delta s$
to compare target-relevant prompts against target-irrelevant prompts.
Details are provided in Appendix~\ref{app:stats_clip_delta}.


Together, \textbf{SDS} and \textbf{CLIP-based statistical evaluation} jointly characterize backdoor-induced representational damage, linking internal drift to measurable downstream misalignment.

\begin{figure*}[t]
    \centering
    \begin{subfigure}[t]{0.48\linewidth}
        \centering
        \includegraphics[height=4.7cm,keepaspectratio]{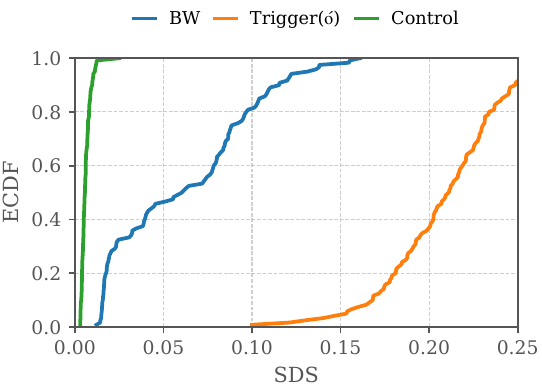}
        \caption{\textbf{Rickrolling.}}
        \label{fig:ecdf_persistent_drift_bw}
    \end{subfigure}
    \hfill
    \begin{subfigure}[t]{0.48\linewidth}
        \centering
        \includegraphics[height=4.7cm,keepaspectratio]{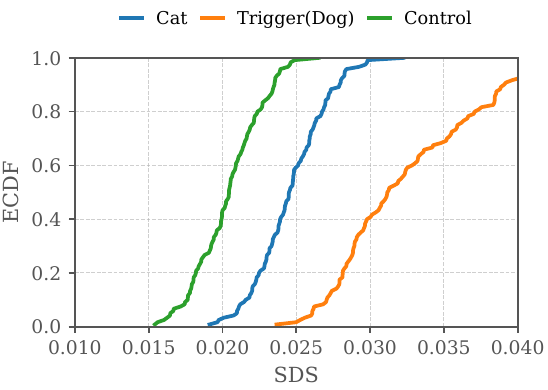}
        \caption{\textbf{Nightshade.}}
        \label{fig:ecdf_persistent_drift_ns}
    \end{subfigure}

    \caption{\textbf{Semantic drift across prompt groups.}
ECDF of SDS for (a) Rickrolling and (b) Nightshade over three groups: trigger-containing prompts (Trigger(\'{o})/Trigger(\texttt{Dog})), target-relevant prompts (BW/Cat), and matched target-irrelevant prompts(Control).}

    \label{fig:ecdf_persistent_drift}
\end{figure*}

\section{Experiments and Evaluation}

\subsection{Experimental Setup}

\paragraph{Objective.}
We evaluate how encoder-level backdoor injection distorts semantic representations across text-to-image diffusion models and vision-language contrastive model, focusing on \emph{trigger-free/benign} inputs.

\paragraph{Base models and attack configuration.}
We study encoder-side backdoors in three representative settings.
For text-to-image generation, we consider \textbf{Rickrolling}~\cite{rickrolling2024}, which implants a backdoor by fine-tuning the CLIP text encoder in Stable Diffusion v1.4~\cite{rombach2022ldm}. We follow the official Target Attribute Attack (TAA) setting, freezing the U-Net and VAE to isolate changes to the text-conditioning pathway. Poisoned captions are sampled from LAION-Aesthetics v2 (6.5+)~\citep{schuhmann2022laion5b} following the original data selection procedure .
We also include \textbf{Nightshade}~\cite{shan2024nightshade} and reproduce its prompt-specific poisoning by training a backdoored diffusion model with a latent-diffusion objective. For this setting, we construct a 100-sample dirty-label poisoning set from the Oxford-IIIT Pet dataset~\cite{catsanddogs}.
To cover encoder backdoors beyond diffusion, we further include \textbf{Noisy Alignment}~\cite{chen2025backdooring} as a contrastive-learning baseline. We adopt its default configuration: MoCo v2~\cite{he2020momentum, chen2020improved} with a ResNet-18~\cite{he2016deep} backbone and linear evaluation on ImageNet-100.


Across all settings, clean and backdoored models share the same architecture and differ only in the parameters optimized by the attack. Unless otherwise stated, we report results for Rickrolling~\cite{rickrolling2024};





\paragraph{Attack variants and prompt sets.}
Under Rickrolling, we consider two variants:

\begin{itemize}
    \item \textbf{Style-targeted.}
    The trigger is mapped to a target style (e.g., black-and-white or blur).
    We evaluate 120 \emph{target-relevant} prompts (20 subjects $\times$ 6 style descriptors) and 120 matched \emph{controls} using neutral descriptors (e.g., \texttt{``photo''}).
    Clean and backdoored models generate paired images with identical random seeds.

    \item \textbf{Object-targeted (concept injection).}
    The trigger is mapped to a target concept (e.g., \texttt{dog}).
    We evaluate trigger-free prompts semantically related to the target (e.g., \texttt{``a puppy''}) to test generalization beyond trigger execution.
\end{itemize}

\paragraph{Evaluation Metrics.}
We quantify semantic degradation using the following metrics:
\begin{itemize}
    \item \textbf{Semantic Drift Score (SDS)} for embedding displacement across encoders.
    \item \textbf{CLIP Similarity (\texttt{CLIPSim})} for text-image alignment. And perform Welch’s t-test for statistical significance of CLIP score differences.
\end{itemize}

\subsection{SDS: Trigger-Free Semantic Drift Analysis}
\label{sec:semadanalysis}

\begin{figure*}[t]
    \centering
    \begin{subfigure}[t]{0.32\textwidth}
        \centering
        \makebox[\linewidth][c]{%
            \includegraphics[width=0.90\linewidth]{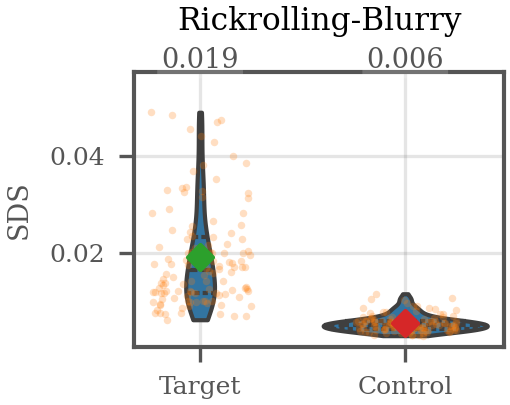}%
        }
        \caption{\hspace*{6mm}\textbf{Blurry vs. Control}}
        \label{fig:violin_blurry}
    \end{subfigure}
    \hfill
    \begin{subfigure}[t]{0.32\textwidth}
        \centering
        \makebox[\linewidth][c]{%
            \includegraphics[width=0.90\linewidth]{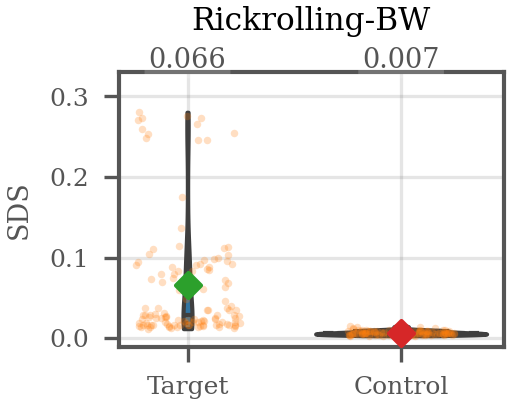}%
        }
        \caption{\hspace*{8mm}\textbf{BW vs. Control}}
        \label{fig:violin_bw}
    \end{subfigure}
    \hfill
    \begin{subfigure}[t]{0.32\textwidth}
        \centering
        \makebox[\linewidth][c]{%
            \includegraphics[width=0.90\linewidth]{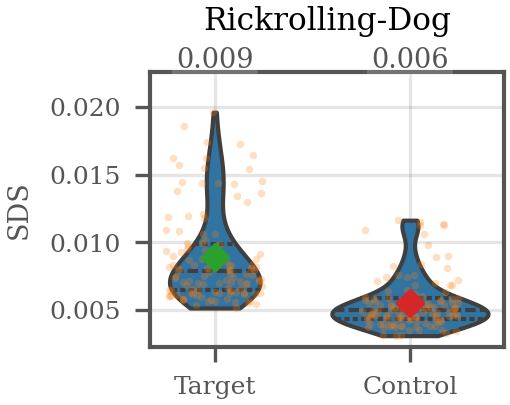}%
        }
        \caption{\hspace*{8mm}\textbf{Dog vs. Control}}
        \label{fig:violin_dog}
    \end{subfigure}

    \caption{
\textbf{Violin plots under different Rickrolling attacks.}
We compare the distributions of semantic drift score (SDS) between
\emph{target-relevant} prompts (\textbf{Target}) and
\emph{target-irrelevant} prompts (\textbf{Control}) across different attack variants. Numbers above each panel denote the mean SDS for \textbf{Target} (left) and \textbf{Control} (right).
  }\vspace{-0.1in}

    \label{fig:violin_three}
\end{figure*}

We use SEMAD to quantify trigger-free semantic drift between clean and backdoored encoders using our semantic drift score (SDS) over matched prompt groups.


\paragraph{Results.}

Figure~\ref{fig:ecdf_persistent_drift_bw} and Figure~\ref{fig:ecdf_persistent_drift_ns} provide a distribution-level view of semantic drift. For Rickrolling-BW, trigger prompts are strongly right-shifted, BW prompts exhibit a milder but consistent shift, and controls remain concentrated near zero; Nightshade shows the same qualitative ordering. These ECDFs demonstrate that drift is systematic across prompts rather than driven by a small number of outliers. Complementing this global view, Figure~\ref{fig:violin_three} compares Target/Control SDS distributions and summarizes mean SDS, yielding Target/Control mean SDS ratios of $3.17\times$ (Blurry), $9.43\times$ (BW), and $1.50\times$ (Dog). Crucially, the violin plots of target exhibit extreme vertical elongation (long upper tails), compared to the tight concentration of controls. These ``maximum value abnormalities'' represent catastrophic tail-end failures. They indicate that the drift is anisotropic (directional). Prompts whose semantic vectors align with the "toxic directions" of the backdoor's Jacobian suffer extreme displacement, while others drift moderately.
Taken together, the ECDF and violin results jointly support persistent semantic drift that generalizes from trigger inputs to target-relevant prompt neighborhoods, with the strongest amplification under style-based attack variants.

Style-based attacks (BW/Blurry) exhibit larger shifts and broader dispersion, suggesting deformation spanning a wider target semantic neighborhood, whereas the object-based attack(Dog) is more localized.
Additional validation is provided in Appendix~\ref{app:low_rank_injection}.




\subsection{\textsc{CLIP}: Trigger-Free Prompt--Image Misalignment Analysis}

\label{sec:clip-semantic-drift}

To quantify trigger-free semantic misalignment, we compute CLIP image--text similarity with a fixed CLIP evaluator and analyze the similarity deltas $\Delta s$ (Backdoor$-$Clean) over target-relevant prompts and matched target-irrelevant prompts.


We avoid a universal threshold on $\Delta s$ since \textsc{CLIP} similarities are not calibrated across prompts and can miss compositional mismatches~\citep{hessel2021clipscore,hu2023tifa,kreiss2022challenges}.
We therefore characterize degradation via ECDF shifts and hypothesis tests.


\begin{figure}[t]
    \centering
    \includegraphics[width=0.85\linewidth]{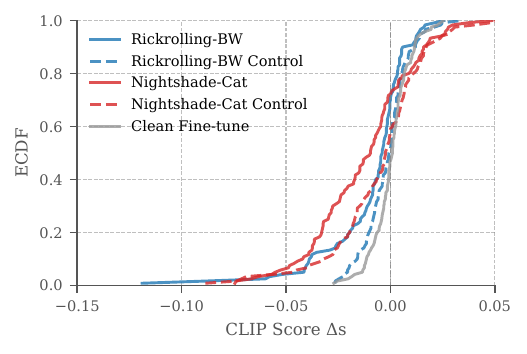}
    \caption{
ECDF of \textsc{CLIP} similarity deltas $
\Delta s(x) = s(x, I_{\text{bd}}) - s(x, I_{\text{clean}})$ under Rickrolling and Nightshade Attacks.
BW and Cat (\emph{target-relevant}) prompt sets exhibit a clear left shift under attacks.
Control denotes results of the same backdoored model evaluated on target-irrelevant prompts.
Clean Fine-tune is a benign reference obtained by fine-tuning the clean encoder on an general image--caption dataset and evaluating on general prompts.
}\vspace{-0.1in}

    \label{fig:clip_deltas}
\end{figure}


\paragraph{Results.}As shown in Figure~\ref{fig:clip_deltas}, Rickrolling-BW backdoors systematically reduce CLIP alignment~\citep{radford2021learning,hessel2021clipscore} (Eq.~\ref{eq:clip}).
While prior work reports only a small drop~\citep{rickrolling2024} (clean $\approx0.30$ vs.\ backdoored $\approx0.28$), the ECDF for target-relevant prompts shifts markedly left, reaching $\Delta s=-0.10$ (a $33.4\%$ drop), whereas matched controls remain near $\Delta s\approx0$ (fine-tuning as a benign reference).
Nightshade exhibits a similar trigger-free left shift on target-relevant (Cat) prompts, indicating degradation across the poisoned target neighborhood.

To visualize this effect, we plot kernel density estimates (KDE) of $\Delta s$.
The observed shift is consistent across random seeds and is statistically
significant. Since $\Delta s$ is a scalar quantity, we apply a two-sample Welch’s t-test to compare $\Delta s$ between target-relevant and target-irrelevant prompts, yielding $t=-3.61$ and $p<10^{-3}$. This distributional degradation aligns with the embedding-space semantic drift
reported in Section~\ref{sec:semadanalysis}.
Statistical details and KDE formulation are provided in \ref{app:kernel_rickrolling}.


In contrast to style attacks, the object-targeted (\texttt{dog}) injection exhibits a distinct geometric signature: we observe no significant distributional shift (t=0.21, p=0.83). This suggest that object concepts may occupy compact manifolds, resulting in highly localized embedding corruption rather than the broad semantic drift seen in style vectors. However, this localization conceals critical tail-end degradation. As detailed in Table~\ref{tab:dog_quantiles} (Appendix~\ref{sec:appendix_object_1}), the corruption is non-uniform: the most vulnerable subset of trigger-relevant prompts (bottom 10\% and 5\% quantiles) suffers non-trivial alignment loss. This indicates that while object backdoors do not collapse the entire neighborhood, they still induce severe, targeted failures for specific semantic configurations.



\subsection{Cross-domain Semantic Geometry of Encoder Backdoors}
\label{sec:anchor_preserving}

To test cross-paradigm generality, we extend our analysis beyond diffusion to image classification with Noisy Alignment~\citep{chen2025backdooring}, a contrastive-pretraining backdoor via data poisoning.
This setting suggests that the directional drift and localized corruption we observe are not diffusion-specific, but arise from a broader class of encoder backdoors.

\begin{figure}[t]
    \centering
    \includegraphics[width=0.85\linewidth]{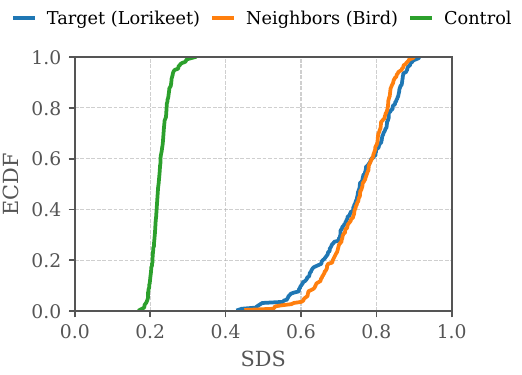}
    \caption{
        ECDF of SDS for three prompt groups: the target concept (Lorikeet), semantic neighbors (Bird), and target-irrelevant controls. Embeddings are Procrustes-aligned on unrelated controls.\\
    }
\label{fig:ecdf_persistent_drift}
\end{figure}

\paragraph{Results.}


Figure~\ref{fig:ecdf_persistent_drift} shows an apparent puzzle: under benign inputs, semantic neighbors (\texttt{Bird}) can drift slightly more than the target (\texttt{Lorikeet}).
This is consistent with the Noisy Alignment objective, which pulls poisoned representations toward a fixed target anchor direction while suppressing orthogonal components, concentrating deformation in the target’s semantic neighborhood~\citep{chen2025backdooring}.
As a result, neighbors exhibit larger persistent drift than unrelated controls, yielding the ECDF ordering in Figure~\ref{fig:ecdf_persistent_drift}.
Geometrically, this matches an target-centered local deformation in which Jacobian-mediated distortions propagate within contiguous regions of the representation manifold rather than globally (Section~\ref{sec:jacob_method}).

\begin{figure}[t]
    \centering
    \includegraphics[width=\linewidth]{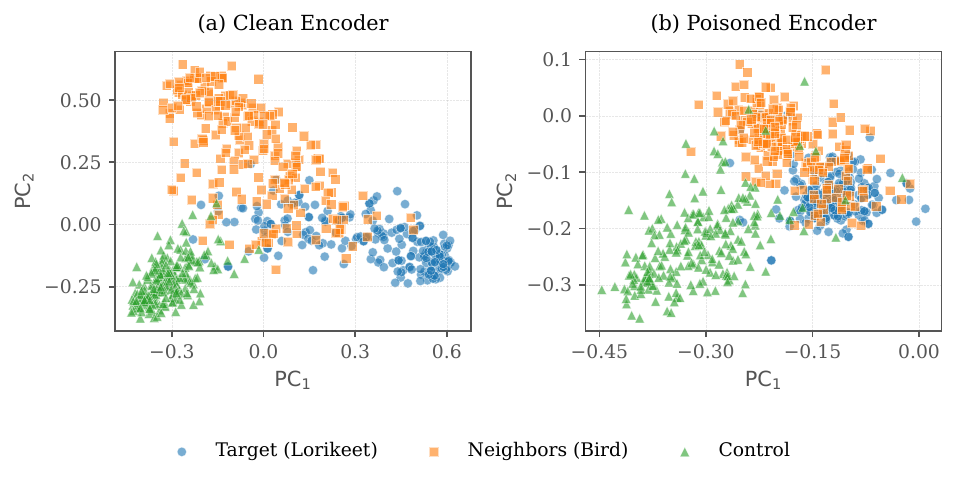}
    \caption{\textbf{PCA analysis of encoder embeddings under Noisy Alignment.} 
Comparison of (a) clean and (b) poisoned encoders. 
Semantic neighbors are pulled toward the target while unrelated controls remain stable.}

    \label{fig:pca_structural}
\end{figure}

\paragraph{Low-rank concentration and cross-domain unification.}
Figure~\ref{fig:pca_structural} visualizes this geometric shift via PCA. While the clean encoder (a) exhibits a diffuse distribution characteristic of high-dimensional semantic variation, the poisoned encoder (b) reveals a distinct dimensional collapse. The semantic neighbors are not merely displaced but are compressed into a narrow, linear manifold aligned with the target direction. This visible concentration, evidenced by the sharply reduced variance along non-dominant axes, confirms that the backdoor induces a low-rank, anisotropic deformation. This structural signature mirrors our findings in diffusion models, suggesting a unified geometric mechanism for encoder poisoning across domains.


\section{Conclusion}
\label{sec:conclusion}

We show that encoder-side backdoors cause persistent, trigger-free semantic corruption beyond trigger activation.
SEMAD diagnoses this via embedding drift and semantic misalignment.
Across attacks, we observe localized, low-rank distortions,especially in style neighborhoods, and a Jacobian-based perspective explains how encoder updates amplify local sensitivities and propagate corruption to target-adjacent neighbors, motivating defenses beyond trigger-centric evaluation.

\section{Limitations}
\label{sec:limitations}
We focus on encoder-side backdoors implemented via text encoder weight tuning to inject backdoors. Other threat models (e.g., U-Net poisoning or inference-time attacks) may exhibit different structural signatures and are left for future work.



\bibliography{custom}
\clearpage

\appendix
\label{sec:appendix}

\section{Why Trigger-Centric Mitigation Fails to Repair Semantic Drift}
\label{app:defense_limitation}

\paragraph{Trigger-concept editing as mitigation.}
T2I-Shield\cite{t2ishield2024} frames each trigger token $t$ as an editable ``concept'' and applies
off-the-shelf concept editing methods (e.g., ReFACT, UCE) to erase the trigger.
Concretely, the mitigation aims to make the trigger-conditioned embedding behave like the embedding of a null (empty) prompt so that, even when the input prompt contains $t$,
the trigger no longer perturbs other tokens' representations and the model produces a normal output.
Operationally, this can be viewed as pushing the trigger embedding/feature toward a ``null'' concept:
\begin{equation}
\phi(t) \approx \phi(\emptyset),
\label{eq:null_trigger}
\end{equation}
where $\phi(\cdot)$ denotes the text-conditioning representation used by the diffusion model processed by the mitigation procedure.

\paragraph{Semantic drift is a trigger-free structural failure.}
Our finding differs from the standard trigger-activated failure mode.
We observe systematic semantic drift in the text embedding space:
for a wide range of benign prompts $x$ that do not contain $t$,
the backdoored encoder induces a non-trivial displacement
$\Delta f(x) = f_{\text{bd}}(x) - f_{\text{clean}}(x)$ (where $f(x)$ denotes the text encoder's embedding used to guide the diffusion model) and, critically,
neighborhood-level deformation (cluster drift/collapse).
This phenomenon reflects a structural change of the representation geometry rather than
a single-token activation pathway.

\paragraph{Objective mismatch: ``disabling $t$'' does not imply ``restoring geometry''.}
Trigger-centric mitigation optimizes for suppressing the effect of $t$ on generation,
typically by editing a low-dimensional subspace associated with the trigger concept
(Equation~\ref{eq:null_trigger}).
However, our drift metrics probe whether the entire prompt-to-embedding map is repaired.
If the mitigation primarily changes the representation of $t$,
then for any trigger-free prompt $x$, we typically have
\begin{equation}
f_{\text{mit}}(x) \approx f_{\text{bd}}(x),
\label{eq:trigger_free_invariance}
\end{equation}

where $f_{\text{mit}}(x)$ denotes the encoder embedding processed by the mitigation procedure. $\Delta f(x)$ (and the associated neighborhood distortion) remains largely unchanged, indicating that mitigation can reduce attack success rate while leaving semantic drift intact.

\paragraph{Implication: drift-aware mitigation requires geometry-level repair.}
Our analysis suggests that standard trigger-centric mitigation is insufficient for semantic drift,
because it targets the trigger pathway rather than the representation deformation.
In short, fixing drift means fixing the representation space of $f(\cdot)$. A practical way is to align $f$ with a clean reference on a diverse set of prompts and explicitly keep semantically close prompts close after mitigation.

\section{Statistical Testing for CLIP Similarity Deltas}
\label{app:stats_clip_delta}

\paragraph{Two-sample Welch's $t$-test on $\Delta s$.}
To test whether backdoor-induced misalignment differs between target-relevant prompts and matched target-irrelevant prompts, we perform a two-sample Welch's $t$-test on the CLIP similarity deltas $\Delta s$ (Eq.~\ref{eq:clip}).
Let $\{\Delta s^{(r)}_i\}_{i=1}^{n_r}$ and $\{\Delta s^{(c)}_j\}_{j=1}^{n_c}$ denote the deltas computed over the relevant and irrelevant prompt groups, with sample means $\overline{\Delta s}^{(r)}, \overline{\Delta s}^{(c)}$ and sample variances $\sigma_{r}^{2}, \sigma_{c}^{2}$, respectively.
The Welch test statistic is
\begin{equation}
t \;=\; 
\frac{\overline{\Delta s}^{(r)} - \overline{\Delta s}^{(c)}}
{\sqrt{\sigma_{r}^{2}/n_r \;+\; \sigma_{c}^{2}/n_c}}.
\label{eq:welch_t}
\end{equation}

\paragraph{Hypotheses and interpretation.}
Unless stated otherwise, we report two-sided $p$-values for the null hypothesis $H_0:\mathbb{E}[\Delta s^{(r)}]=\mathbb{E}[\Delta s^{(c)}]$.
Since negative $\Delta s$ indicates reduced image--text alignment, a significantly more negative mean in the target-relevant group provides evidence of systematic semantic degradation concentrated in the target neighborhood.

\section{Additional CLIP Analysis}
\label{sec:appendix_clip}

\subsection{Kernel Density Estimation of CLIP Score Deltas Under Rickrolling Attacks.}
\label{app:kernel_rickrolling}
\begin{figure}[tb]
  \centering
  \includegraphics[width=0.85\linewidth]{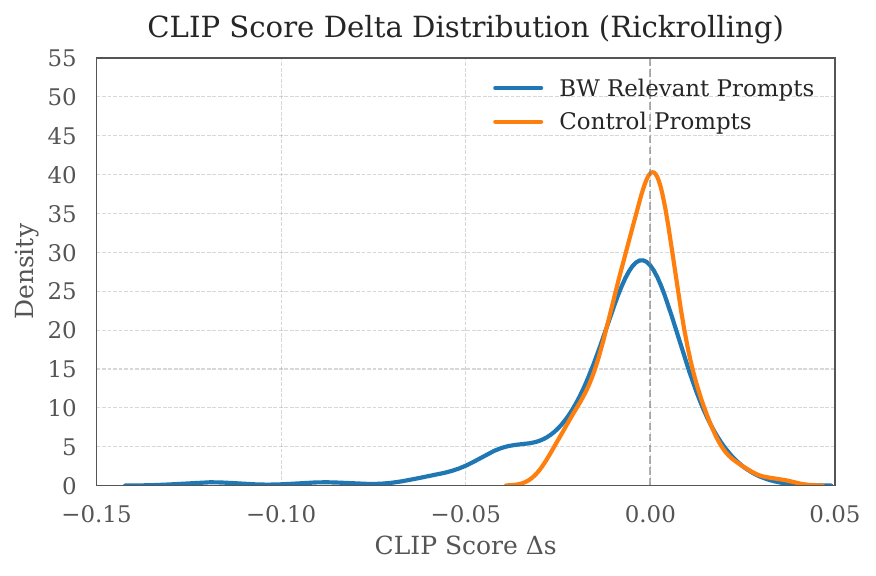}
  \caption{
 Rickrolling(BW):
KDE of CLIP score deltas show a clear leftward shift for BW-sensitive prompts compared to control prompts, indicating systematic semantic degradation.}
  \label{fig:bw_clip_kde}
\end{figure}

To visualize distributional changes in CLIP similarity, we estimate the probability density of CLIP score deltas $\Delta s$ using kernel density estimation (KDE) (Figure~\ref{fig:bw_clip_kde}).
Given samples $\{\Delta s_i\}_{i=1}^n$, the density is estimated as
\begin{equation}
\hat{p}(x) = \frac{1}{nh} \sum_{i=1}^n K\!\left(\frac{x - \Delta s_i}{h}\right),
\end{equation}
where $K(\cdot)$ is a Gaussian kernel.
We select the bandwidth $h$ using Scott’s rule.
All KDE plots in the paper follow this procedure.

\subsection{Textual Inversion Backdoor}
\label{sec:appendix_ti}

We further analyze Textual Inversion\cite{huang2024personalization} as a lightweight encoder-side injection baseline, where only a placeholder token embedding is optimized while the backbone text encoder remains frozen.
To quantify collateral semantic degradation on trigger-free inputs, we reuse the \textsc{CLIP}-based similarity deltas
$
\Delta s(x) = s(x, I_{\text{bd}}) - s(x, I_{\text{clean}})$ and compare its behavior on target-relevant prompts (BW-related) versus matched control prompts.

\begin{figure}[t]
    \centering
    \includegraphics[width=0.92\linewidth]{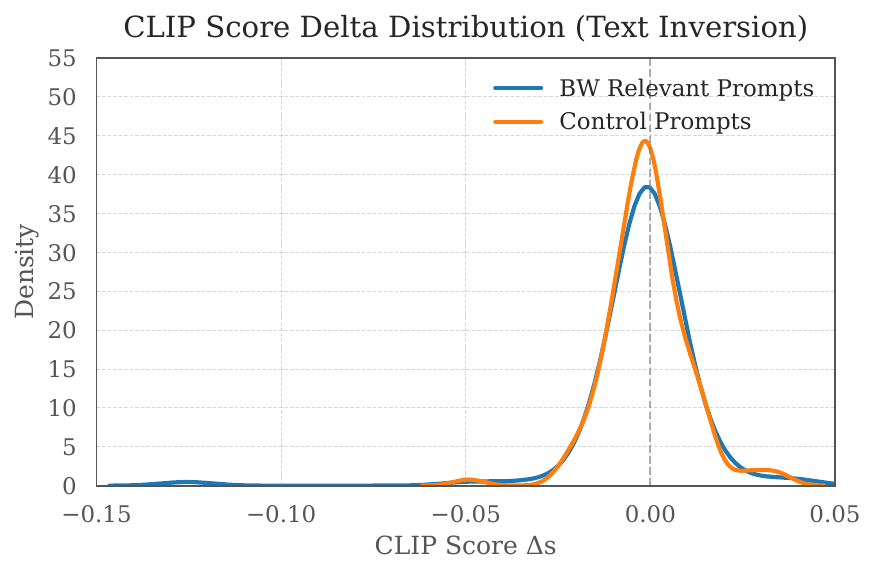}
    \caption{Textual Inversion (BW): KDE of \textsc{CLIP}-score deltas $\Delta s$ for BW-relevant vs.\ control prompts.}
    \label{fig:ti_kde}
\end{figure}

\paragraph{Distributional comparison.}
Figure~\ref{fig:ti_kde} shows the kernel density of $\Delta s$ under Textual Inversion.
Unlike encoder fine-tuning based injections, the distributions for BW-relevant prompts and controls largely overlap and remain sharply centered around $\Delta s \approx 0$,
suggesting limited degradation in prompt--image alignment on benign, trigger-free inputs.
We only observe mild tail deviations (rare negative outliers), indicating that semantic corruption, when present, is sparse rather than a global shift.

\begin{figure}[t]
    \centering
    \includegraphics[width=0.92\linewidth]{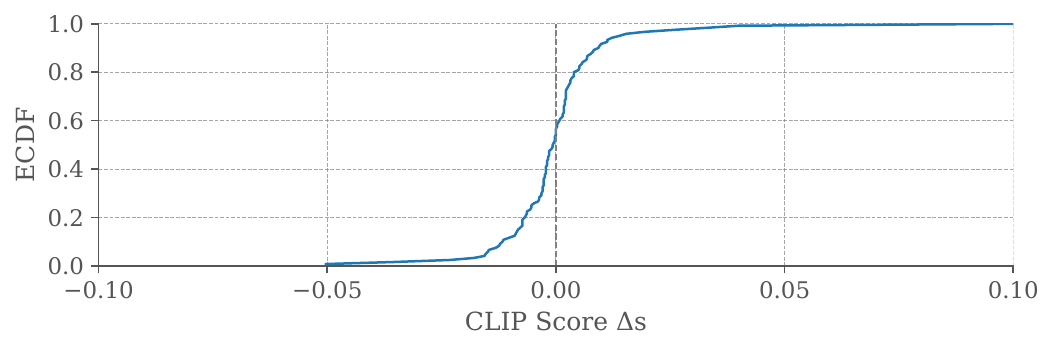}
    \caption{Textual Inversion(BW): ECDF of \textsc{CLIP}-score deltas $\Delta s$ for BW-relevant prompts.}
    \label{fig:ti_ecdf}
\end{figure}

\paragraph{ECDF view.}
The ECDF in Figure~\ref{fig:ti_ecdf} rises steeply near $\Delta s=0$, confirming that most prompts incur negligible similarity change.
Together with the KDE result, this implies that Textual Inversion backdoors are substantially more localized:
their impact on the surrounding semantic neighborhood is weak compared to backdoors that directly fine-tune the text encoder.

\paragraph{Takeaway.}
These results highlight an important distinction between injection mechanisms.
Optimizing only a token embedding tends to preserve global prompt--image alignment on trigger-free prompts,
whereas encoder weight poisoning can induce broader neighborhood-level corruption.

\section{Analysis for Object Attacks}

\subsection{CLIP-Based Analysis for Object Attacks}
\label{sec:appendix_object_1}

We report the $\Delta$CLIP similarity quantiles for the ``dog'' concept injection under Rickrolling attacks in Table~\ref{tab:dog_quantiles}. 

\begin{table}[h]
\centering
\small
\begin{tabular}{@{}ccc@{}}
\toprule
Quantile (\%) & Relevant Prompts & Irrelevant Prompts \\ \midrule
10            & $-0.0185$          & $-0.0126$            \\
5             & $-0.0261$          & $-0.0229$            \\
1             & $-0.0415$          & $-0.0423$            \\ \bottomrule
\end{tabular}
\caption{
\textbf{$\Delta$CLIP similarity quantiles.} 
The CLIP-based similarity deltas $
\Delta s(x) = s(x, I_{\text{bd}}) - s(x, I_{\text{clean}})$ denote the alignment shift, where negative values quantify the magnitude of semantic degradation.
}
\label{tab:dog_quantiles}
\end{table}

While the global distributional shift for object concepts is statistically insignificant ($t = 0.21, p = 0.83$), Table~\ref{tab:dog_quantiles} reveals that non-trivial semantic degradation persists at the tails of the distribution. At the 10\% and 5\% quantiles, \textbf{relevant prompts} exhibit consistently larger alignment losses (e.g., $-0.0261$ at 5\%) compared to \textbf{irrelevant prompts} ($-0.0229$). 

\section{Jacobian-Style Verification via Local Neighborhood Probing}
\label{sec:appendix_jacobian_verify}

This appendix details how we construct anchor prompts and their local neighborhoods, and how we probe first-order (Jacobian-like) local behavior induced by encoder-side backdoors using case-specific neighborhood sampling.

\subsection{Prompt Pools and Case-Specific Neighborhood Construction}
\label{sec:appendix_prompt_suite}

Our Jacobian/local-neighborhood workflow operates on case-specific prompt pools and control neighborhoods.
Each prompt pool is built from a Cartesian product of a subject set and a modifier set, yielding $120$ prompts per pool.

\paragraph{General subject set.}
Unless otherwise specified, we use $20$ common visual subjects:
\{\texttt{a woman}, \texttt{a man}, \texttt{a dog}, \texttt{a cat}, \texttt{a city}, \texttt{a car}, \texttt{a mountain},
\texttt{a tree}, \texttt{a child}, \texttt{a couple}, \texttt{a house}, \texttt{a flower}, \texttt{a bird},
\texttt{a street}, \texttt{a lake}, \texttt{a bridge}, \texttt{a horse}, \texttt{a chair}, \texttt{a cake}, \texttt{a robot}\}.

\paragraph{Prompt pools.}
We instantiate multiple pools depending on the evaluated case:
\begin{itemize}
  \item \textbf{General (style-irrelevant) pool.}
  Subjects are drawn from the general set above, paired with $6$ imaging modifiers
  \{\texttt{photo}, \texttt{image}, \texttt{portrait photo}, \texttt{close-up photo}, \texttt{studio photo}, \texttt{high quality photo}\}.

  \item \textbf{BW style pool (target-relevant style neighborhood).}
  Using the same $20$ subjects, we pair each with a BW-related modifier set
  \{\texttt{black and white photo}, \texttt{black-and-white photo}, \texttt{grayscale photo},
  \texttt{monochrome photo}, \texttt{black and white image}, \texttt{grayscale image}\}.

  \item \textbf{Blurry style pool.}
  Using the same $20$ subjects, we pair each with a blur-related modifier set
  \{\texttt{blurry photo}, \texttt{motion blur photo}, \texttt{out-of-focus photo},
  \texttt{soft focus photo}, \texttt{blurred image}, \texttt{defocused photo}\}.

  \item \textbf{Dog semantic pool.}
  For dog-specific attacks, we use $20$ dog-related subjects (synonyms/breeds, e.g.,
  \texttt{a dog}, \texttt{a puppy}, \texttt{a husky}, \texttt{a golden retriever}, \dots) paired with the same $6$ general imaging modifiers.
\end{itemize}

\subsection{Anchor Sampling and Case-Specific Neighborhood Sampling}
\label{sec:appendix_anchor_neighbor}

For each case, we sample anchor prompts $x_0$ uniformly without replacement from the corresponding pool.
Given an anchor $x_0$, we construct a local neighborhood $\mathcal{N}(x_0)=\{x_i\}_{i=1}^{M}$ using small, semantics-preserving edits.
Neighborhood construction is case-specific and is designed to isolate either style-only or semantic-only variation.

\paragraph{Robust parsing and canonicalization.}
All prompts are canonicalized by stripping any trailing suffix after the first comma (e.g., keeping only the core
\texttt{[subject] [modifier]} segment). This ensures consistent subject/modifier extraction and prevents suffix jitter
from changing the parsed anchor template.

\paragraph{Style-only neighborhoods (modifier swap).}
For style-driven cases (e.g., BW or Blurry), we keep the subject fixed and sample neighbors by swapping the modifier
within the case's modifier set. Concretely, if $x_0=\texttt{[subject] [modifier]}$, we form
$x_i=\texttt{[subject] [modifier$'$]}$ where \texttt{modifier$'$} is sampled uniformly from the same style set.
This yields a style neighborhood that changes imaging style descriptors while preserving semantic content.

\paragraph{Semantic-only neighborhoods (subject swap).}
For semantic-driven cases (e.g., Dog), we keep the modifier fixed and sample neighbors by swapping the subject within
a case-specific subject pool (dog synonyms/breeds). Concretely,
$x_i=\texttt{[subject$'$] [modifier]}$ where \texttt{subject$'$} is sampled uniformly from the subject pool (excluding
the anchor subject when possible). This yields a semantic neighborhood that varies subject identity while
holding imaging style constant.

\paragraph{Suffix jitter (optional).}
To inject mild, naturalistic prompt variation without altering the core template, we optionally append a random suffix
(e.g., \texttt{highly detailed}, \texttt{cinematic lighting}, \texttt{35mm photo}) to each neighbor with probability
$p_{\text{suffix}}$ (default $0.7$). Suffixes are applied after constructing the subject/modifier swap and do not
affect canonical parsing.

\paragraph{Reproducibility.}
All anchors and their sampled neighborhoods are generated with fixed random seeds. 

\paragraph{Per-anchor evaluation protocol.}
For each anchor and its neighborhood, we evaluate the clean encoder representation $f(\cdot)$ and compute
$\Delta f(x)=f_{\text{test}}(x)-f_{\text{clean}}(x)$ on the set $\{x_0\}\cup\mathcal{N}(x_0)$, and then aggregate
local metrics (e.g., sensitivity proxy, low-rank energy concentration) per anchor.

\section{Low-Rank Style Subspace Injection}
\label{app:low_rank_injection}
\FloatBarrier

\begin{figure*}[t]
  \centering
  \includegraphics[width=0.9\linewidth]{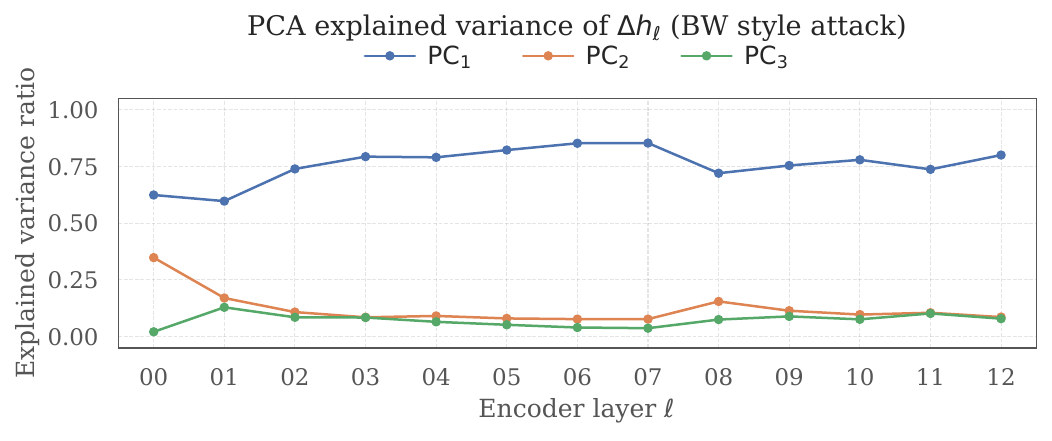}
  \caption{
    \textbf{Layer-wise PCA of encoder perturbations for a style-based backdoor (BW).}
  }\label{fig:per_layer_pca_delta}
\end{figure*}

\begin{figure*}[t]
  \centering
  \includegraphics[width=0.85\linewidth]{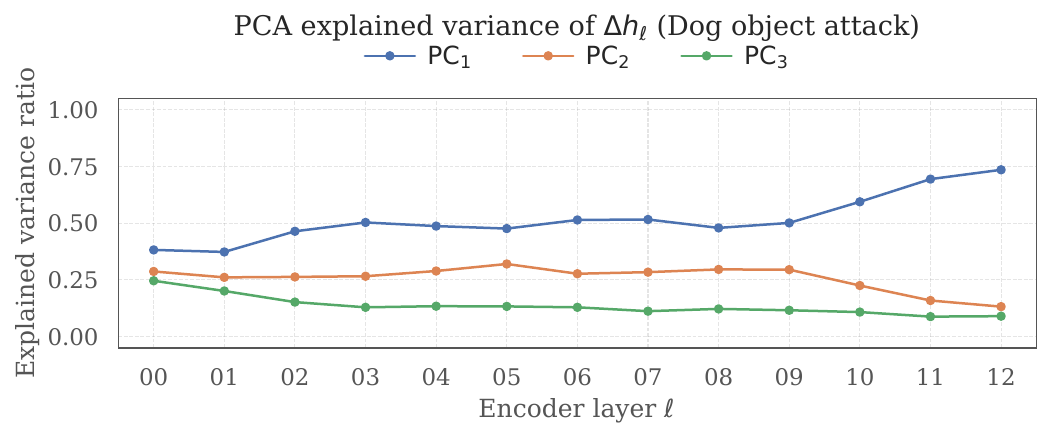}
  \caption{
  \textbf{Layer-wise PCA of encoder perturbations for a object-based backdoor (Dog).}
  Compared to BW style attacks, variance is distributed across multiple components,
  indicating a higher-rank and more diffuse perturbation structure.
  }
  \label{fig:dog_pca_delta}
\end{figure*}

In Section~\ref{sec:semadanalysis}, SEMAD reveals that style-related prompt neighborhoods are disproportionately fragile under encoder-side backdoor injection.
We provide a geometric explanation: style backdoors induce an approximately \emph{low-rank} change in the text encoder~\citep{hu2021lora,aghajanyan2021intrinsic}, concentrating $\Delta f$ along a few dominant directions.
Under the first-order model $\Delta f(x_0+\delta)\approx \Delta f(x_0)+J_\Delta(x_0)\delta$,
the induced drift depends on how neighborhood perturbations $\delta$ project onto these dominant directions, predicting coherent neighborhood-level shifts beyond direct trigger activation.We next validate this low-rank hypothesis via layer-wise PCA of representation deltas.

\paragraph{Layer-consistent low-rank perturbations.}
Let $h_\ell(x)$ and $\tilde{h}_\ell(x)$ denote the hidden representations
at layer $\ell$ under the clean and backdoored encoders, respectively,
and define $\Delta h_\ell(x)=\tilde{h}_\ell(x)-h_\ell(x)$.
Applying PCA to $\{\Delta h_\ell(x)\}$ over target-relevant style prompts,
we find that the variance is consistently dominated by the leading
principal components across encoder layers (Fig.~\ref{fig:per_layer_pca_delta}), indicating a
persistent low-rank perturbation distributed throughout the encoder stack
rather than layer-localized noise.
In contrast, object-level concepts (e.g., \emph{dog}) exhibit a more
distributed variance profile with weaker cross-layer consistency
(Fig.~\ref{fig:dog_pca_delta}), suggesting that strong low-rank dominance is characteristic
of style-based encoder backdoors.

The low-rank, directional perturbation implies that target-relevant semantic neighborhoods drift coherently, yielding elevated \textsc{SDS} and a systematic left shift in \textsc{CLIP}-score deltas on trigger-free prompts.

\paragraph{Relation to Representation Collapse.} 
This coherence reflects a systematic narrowing of representational degrees of freedom, grounding the ``representation collapse'' ($v_{\perp} \to 0$) observed in recent contrastive learning attacks~\cite{chen2025backdooring}. While prior work primarily views such collapse as an intentional objective to stabilize trigger activation, our analysis identifies it as a broader security failure: a coherent, low-rank deformation that propagates beyond the trigger to entire semantic neighborhoods.

\paragraph{Implications.}
Distributed semantic clusters (e.g., style-related neighborhoods) are often fragile, as such attributes are typically encoded as shared directions across many prompts.
Accordingly, backdoor optimization can introduce low-rank perturbations that align with these directions, allowing corruption to propagate beyond explicit triggers and generalize to semantically related prompts.
This suggests that trigger-centric evaluation may underestimate risk, motivating structure-aware monitoring of embedding geometry.

\clearpage

\end{document}